# Room-temperature transverse-electric polarized intersubband electroluminescence from InAs/AlInAs quantum dashes

V. Liverini[1,a)], L. Nevou[1], F. Castellano[1,2], A. Bismuto[1], M. Beck[1], Fabian Gramm[3] and J. Faist[1]

[1] ETH Zürich, Institute for Quantum Electronics, Wolfgang-Pauli-Strasse 16, 8093 Zürich, Switzerland

[1,2] CNR, Istituto Nazionale, Piazza dei Cavalieri 12, Pisa, Italy

[3] Electron Microscopy ETH Zurich (EMEZ), ETH Zurich, Wolfgang-Pauli-Strasse 16, 8093 Zürich, Switzerland

## Abstract

We report the observation of transverse electric polarized electroluminescence from InAs/AlInAs quantum dash quantum cascade structures up to room temperature. The emission is attributed to the electric field confined along the shortest lateral dimension of the dashes, as confirmed by its dependence on crystallographic orientation both in absorption measurements on a dedicated sample and from electroluminescence itself. From the absorption we estimate a dipole moment for the observed transition of <x>=1.7 nm. The electroluminescence is peaked at around 110 meV and increases with applied bias. Its temperature dependence shows a decrease at higher temperatures limited by optical phonon emission.

Since their first demonstration in 1994, quantum cascade lasers (QCLs)[1] have shown to be efficient sources of coherent light in the mid-infrared spectral region and have extended their coverage also to the THz region. Their operation relies on intersubband transitions in quantum wells (QWs), grown epitaxially in the active region. The lack of in-plane confinement of QWs exposes the electrons to numerous in-plane scattering paths, which decrease the upper laser state lifetime of the laser, causing an increase in the threshold current and a decrease of the wallplug efficiency. Despite this drawback, tremendous improvements in active region design, waveguide engineering and processing have been achieved in the mid-IR region[2,3], while the THz region is still suffering from poorer performances and a lack of room-temperature operation. Moreover, due to intersubband selection rules[4], light propagation parallel to the growth direction ($z$) is forbidden for both polarization, transverse electric (TE or $s$) and transverse magnetic (TM or $p$), making surface emission impossible. On the other hand, light propagation in the direction perpendicular to the z-direction is allowed but only for TM-polarized light, i.e. with the electric field component along the z-direction, where the electrons are confined by the QWs. Indeed light from QCLs is TM-polarized and surface emission is achieved only by using special out-coupling schemes[5]. In order to increase the upper laser state lifetime and allow for TE-polarized light in QCLs, it was proposed to substitute the QWs in the active region by 3D confined quantum dots (QDs)[6,7]. Since the active region of QCLs requires the growth of hundreds of layers of high optical and electronic quality, self-assembled techniques for the growth of the QDs are preferred. The well-studied InAs QDs nucleated by Stranski-Krastanov growth mode on GaAs have been used originally to demonstrate mid-infrared emission due to their easier fabrication compared to other material systems[8,9,10,11]. We recently pointed out that one problem with the GaAs material system is the lack of a suitable extraction mechanism from the levels confined in the QDs[12], which prevents the engineering of an efficient depopulation of the lower laser level without the introduction of undesired strain, which cannot be easily compensated. Therefore, we proposed to use QCLs based on InP, where, by working in the InGaAs/AlInAs lattice-matched system, extraction schemes can be easily engineered. Moreover, in this material system we can benefit from a well-known processing technology, which allows for the use of extremely low-loss waveguides. However, dot formation on the AlInAs/InGaAs compounds is more difficult due to the lower lattice-mismatch between them and InAs. As a result the QDs are usually elongated along the [1-10] orientation, hence the terminology quantum dashes (QDashes), and are highly inhomogeneous[13]. Nonetheless, InAs QDashes have been proven successful for interband lasers[14, 15] and we recently demonstrated a broad emission from quantum cascade devices based on these nanostructures[12, 16]. Here we present the characterization of quantum cascade structures based on InAs/AlInAs QDashes with improved homogeneity. The electroluminescence (EL) of these devices is clearly TE-polarized for light propagating along the elongation direction of the dashes in agreement with intersubband absorption measurements.

The molecular beam epitaxy (MBE) –grown nanostructures at the core of the devices were optimized for low inhomogeneous broadening by growing simple photoluminescence (PL) test samples consisting of InAs self-assembled QDashes embedded in $Al_{0.48}In_{0.52}As$ lattice-matched to the InP substrate. To examine the morphology of the dashes, these test structures were topped by uncapped InAs dashes grown with the same conditions as the embedded ones. An SEM picture of the dashes is shown in the inset of Fig. 1a. Based on the full width at half maximum (FWHM) of the PL at 5K, the most homogeneous QDashes were grown at 460ºC with a nominal InAs coverage of 6 monolayers (MLs). Their PL emission was centered at ~890 meV and the FWHM was ~33 meV for a low incident pump power of ~ 10 W/cm$^2$. As estimated from the inset in Fig.1a, the density of the QDashes was approximately $3\times10^{10}$ cm$^{-2}$ and they were elongated in the [1-10] direction by an average length of ~65 nm, while the width was ~15 nm. For both directions we still notice a large inhomogeneity, indicating that the improvement in PL FWHM is mainly in the height of the dashes, which was estimated to be ~2.5 nm from TEM measurements. An absorption structure containing 50 layers of QDashes was grown on a semi-insulating InP substrate. The dashes were doped at $n_v=9\times10^{17}$ cm$^{-3}$ to populate their ground state and allow for intersubband absorption. Considering the measured QDash density this amounts to about 5-6 electrons per dash. The transmission through the polished substrate was measured at 300K in a Fourier transform infrared spectrometer (FTIR) at normal incidence for linearly polarized light: in one case the sample was rotated so that the electric field would be in the [1-10], or *y,* direction (direction 1 in Fig. 1b), and the other case so that it would be in the [110], or *x,* direction (direction 2 in Fig. 1b). Dividing the measurement taken in direction 2 by the one taken in direction 1 we obtained the spectrum shown in Fig. 1a. A clear absorption dip was observed at room temperature. The intensity of the light after going through the sample was reduced by ~4% ($I/I_0$~0.96). The absorption was fitted with a Gaussian curve centered at 89 meV and with a FWHM of 28.5 meV. By assuming a refractive index $n_{ref} = 3.41$, and a sheet density $n_s = 1.6\times10^{11}$ cm$^{-2}$ (derived from the volumetric sheet density $n_v$ and the nominal thickness of the InAs coverage of 6 MLs), we could obtain a dipole moment for the transition corresponding to <x>. = 1.7 nm. These results are in agreement with a previous study aimed at the realization of quantum dot infrared photodetectors (QDIPs) based on similar nanostructures[17]. Similarly to this study, we can adopt the following notation for the crystallographic directions: *x* is the direction of the width of the dashes, [110], *y* is the direction of the length of the dashes, [1-10], and *z* is the direction of the height of the dashes, [001]. If the state is represented by the notation $|n_x, n_y,$ $n_z>$, then we can denote the observed transition as the transition from the ground state *$e_{111}$*=|1,1,1> to the excited state *$e_{211}$*=|2,1,1>. To compute the band structure of our quantum dashes, we developed a 3 dimensional Schrodinger solver and we modeled an $In_{0.85}Al_{0.15}As$ quantum dash as a semi-cylinder of dimensions: width = 15 nm, length = 65 nm and height = 2.5 nm. We obtained an energy difference $e_{211}$-$e_{111}$ of 120 meV (state wavefunctions shown in the bottom left of Fig. 1b). This transition has a computed dipole moment in the *x*-direction <x>= 2.6 nm, amounting to an oscillator strength $f_x$ = 22. It should be noted that a similar energy difference, dipole moment and oscillator strength can be obtained from the energy levels associated with the $n_y$=2 levels, namely *$e_{121}$*=|1,2,1> and $e_{221}$=|2,2,1> (Fig. 1b, bottom right), and so on for higher levels. Our doping was enough to populate the |1,1,1>, |1,2,1> and |1,3,1>, therefore the absorption can be attributed to transitions starting from these levels to the $n_x$=2 levels.

Furthermore, we observe a strongly *p*-polarized absorption peak at~365 meV (FWHM=65meV), which, in agreement with literature [18,17], is attributed to absorption from the ground state |1,1,1> to either a WL or the $n_z$=2 level, namely $e_{112}$=|1,1,2>. Confirmation of the location of the QDash ground state was also provided by current-voltage measurements on QDash-based double-barrier resonant tunneling diodes.

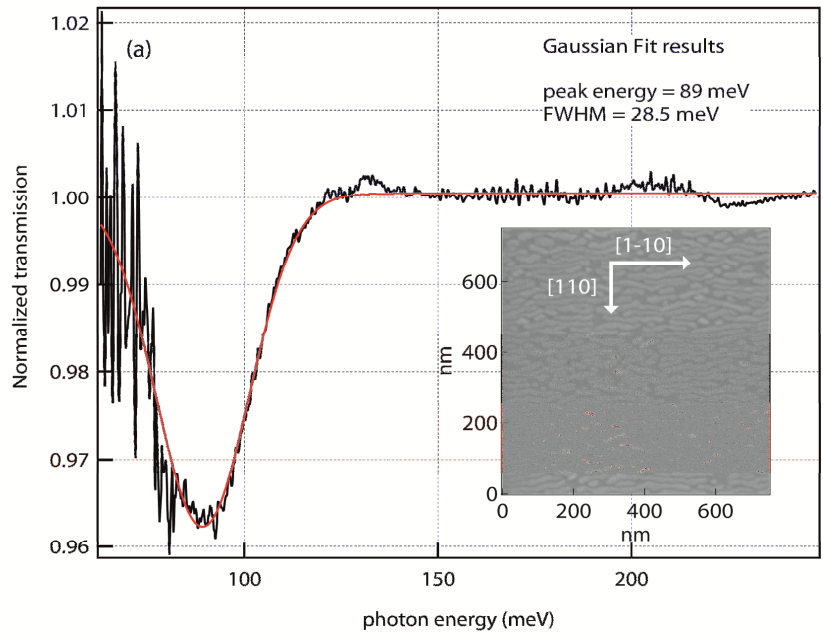

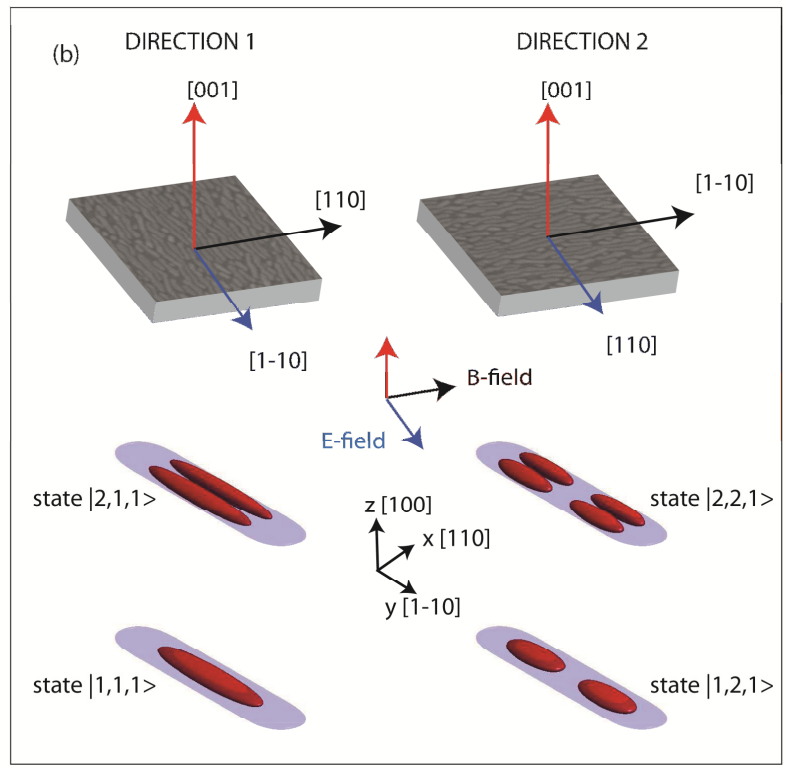

**FIG. 1** (a) Intersubband transmission measurement (black) and fit (red) of the sample consisting of 50 layers of InAs/AlInAs QDashes (T= 300K). Inset: SEM image of the uncapped QDashes. (b) Top: scheme of the normal incidence absorption measurement. Bottom: electron wave functions computed with the 3D Schroedinger solver for the |1,1,1> and the |2,1,1> energy levels (left) and for the |1,2,1> and the |2,2,1> energy levels (right).

Based on the above characterization, we designed a quantum cascade structure, where the transition observed in absorption could be used as a laser transition. We used a self-consistent 1D Schroedinger Poisson solver to design the structure. The active region at an electric field of 76 kV/cm can be seen in Fig. 2a. It consists of a relatively short injection/extraction miniband with a doped region of $n_s$=2.5×$10^{11}$ $cm^{-2}$. We aimed at injecting the electrons in the upper state of the transition measured in absorption, i.e. about 90 meV above the ground state. Therefore, the electrons are injected in the upper state $|2,1,1>$ and they emit light by relaxing to the ground state $|1,1,1>$ from which they are extracted by the following miniband.

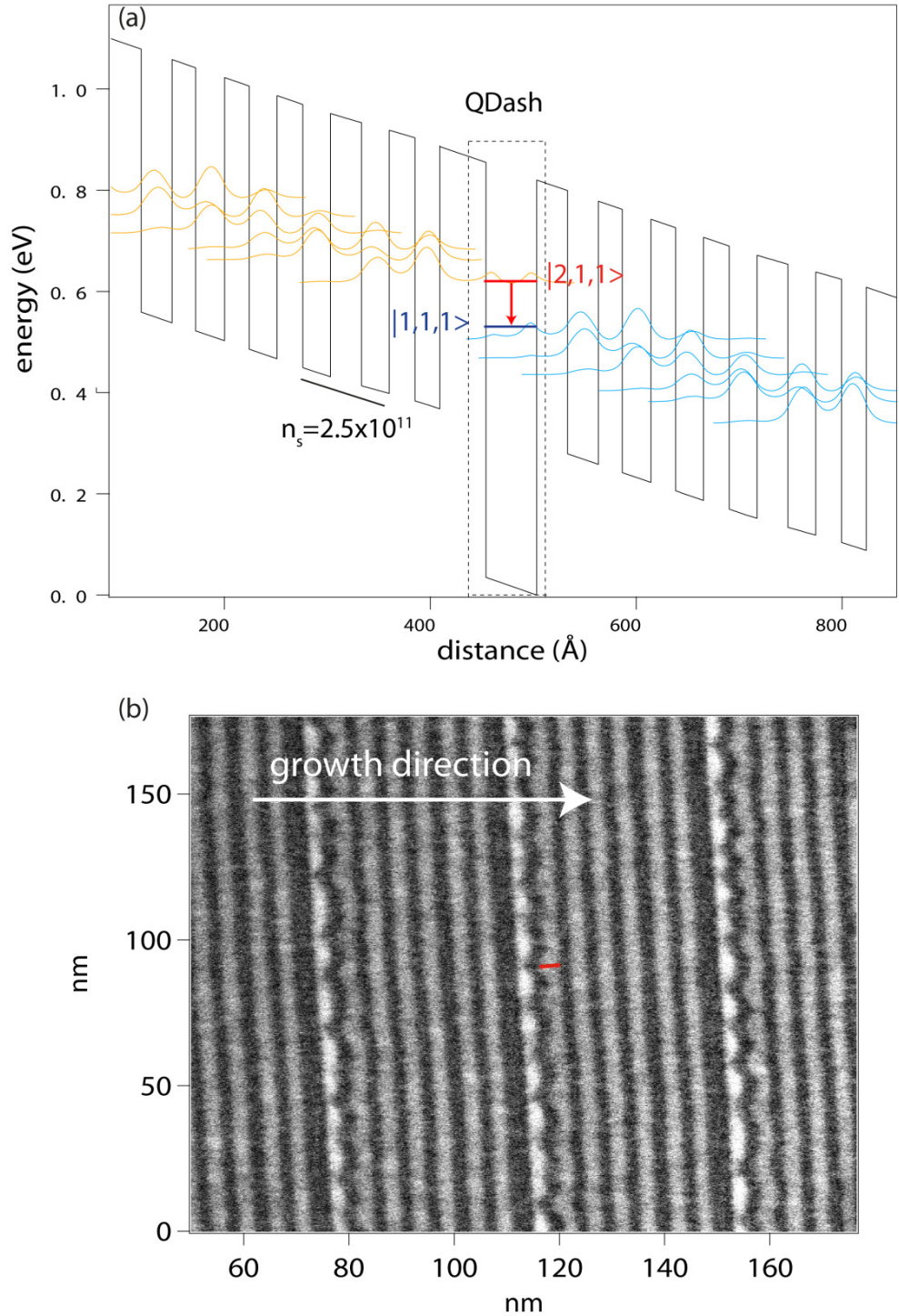

**FIG. 2** (a) Design of the quantum cascade structure at an electric field of F=76 kV/cm. (b) Z-contrast STEM image of three out of the 50 active regions grown. The thick red line indicates the part of the QW, which is much thicker compared to the design.

The active region, which was repeated 50 times, was grown at the temperature optimized for the QDash growth and the dashes were nucleated on top of the injection barrier. A z-contrast STEM image of approximately three active regions can be seen in Fig. 2b. We notice that the layers after the QDash growth are planarized only after the first InGaAs extraction well. This occurs similarly for all the 50 periods without any noticeable difference between the initial and final periods. After MBE growth, a cladding layer based on Si-doped InP was re-grown on top of two quarters of the wafer by metal-organic vapor phase epitaxy (MOVPE). Each quarter was then processed with a standard laser process with relatively wide (30, 40 and 50 µm) ridges. The ridges were wet-etched and isolated using a 350 nm SiNx layer. In one quarter (B) the ridges were parallel to the [110] (*x*) direction, while on the other quarter (D) they were parallel to the [1-10] (*y*) direction. We analyzed two similar devices (ridge width ~30 µm and ridge length ~1.5 mm), one from sample B and one from sample D. The samples were mounted on a continuous flow cryostat and were cooled to a temperature of 80K. The EL measurement was performed while pumping the samples in pulsed mode with a 300 ns pulse at a duty cycle of 3% and at a bias of ~25 V (corresponding to a pump current of 4.5 A). The light was analyzed via an FTIR in step scan mode using a lock-in amplifier triggered by the pulser to increase the signal to noise ratio. To collect the signal we used a liquid $N_2$-cooled mercury cadmium telluride (MCT) detector with a cutoff at ~14 µm. Fig. 3a and 3b show the TE and TM polarized EL from sample B (thin lines) and sample D (thick lines), respectively. Fig. 3a shows the TE-polarized emission. A peak centered at 110 meV with a FWHM of ~52 meV is noticeable only for sample D, where the ridge is along the elongation of the dashes (*y*-direction). The electric field, in this case, is in the confinement direction related to the width of the dashes (*x*-direction). In the case of sample B, the TE polarized transition addressed by the injector cannot be coupled out because the light propagates in the other direction. This peak is associated to the absorption peak at 89 meV shown in Fig. 1a. Since the EL spectral region is very close to the cutoff of the detector, we used a blackbody radiation to calibrate the response of the measuring apparatus. Corrections on the spectrum according to this calibration shifted the peak to ~100 meV, which is in better agreement with the absorption measurements. In Fig. 3b, we show the TM EL, which is, instead, the same for both directions and is therefore associated with the confinement in the growth direction. Specifically, two peaks appear: an intense peak at 233 meV and a smaller peak about 39 meV below. We attribute these peaks to transitions within the first InGaAs QW after the QDash layer. Indeed, it can be observed in the TEM image that after the growth of the QDashes the following InGaAs QW is up to 30% larger than the design in the region between two QDashes (Fig. 2b, thick red line). The TM-polarized EL is about 4.5 times larger than the TE-polarized one, which is expected since the TE signal is at lower spectral frequencies.

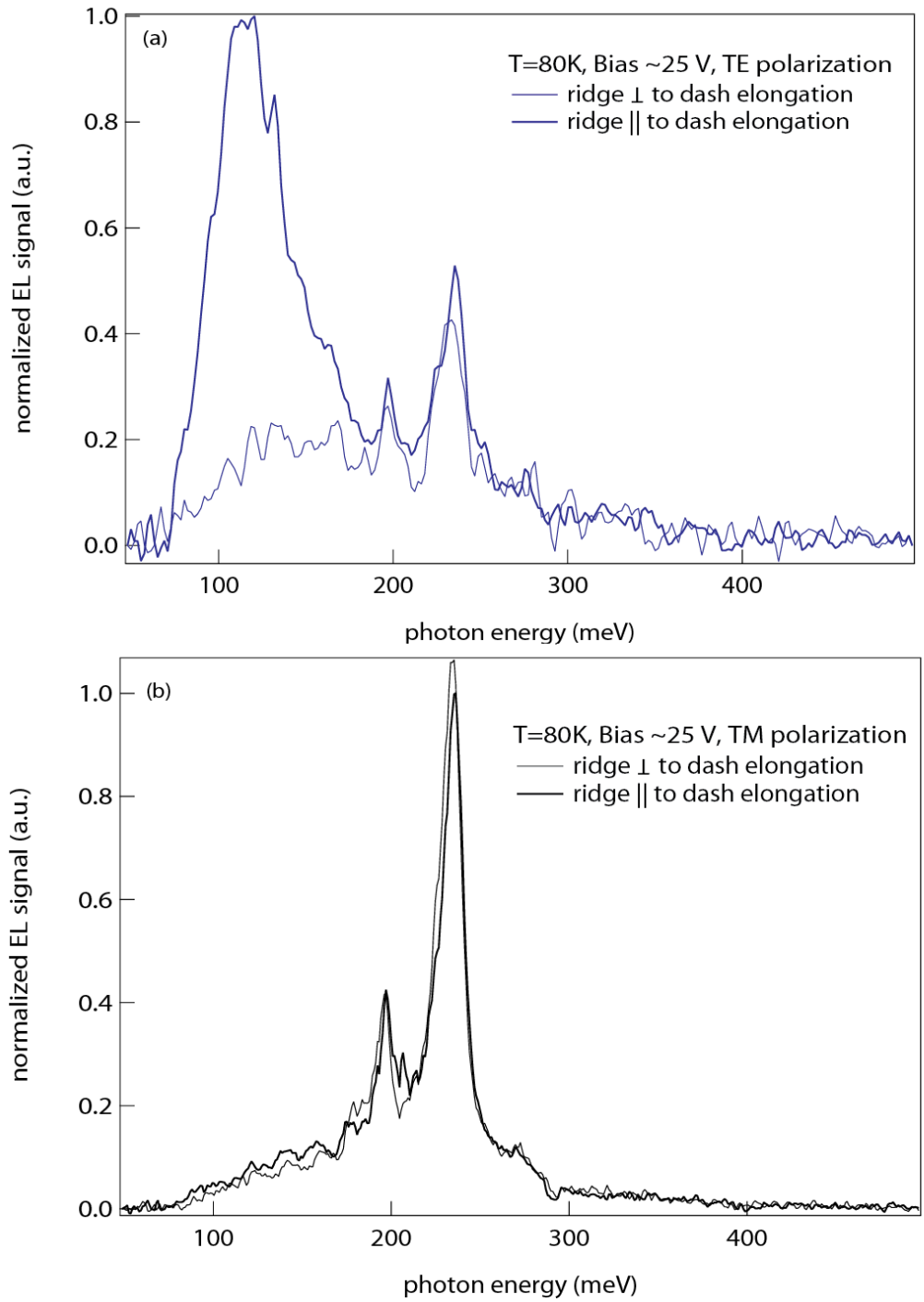

**FIG. 3** a) TE emission from sample D, with the ridge in the same direction as the length of the QDashes, [1-10]-direction (thick line) and from the sample B, with the ridge perpendicular to the [1-10] direction. b) same as above for the TM emission.

In addition, we analyzed the TE-polarized EL from sample D as a function of bias and temperature. The inset in Fig. 4a shows the light-current-voltage (LIV) characteristic curve for the device taken by pumping it with a duty cycle of 3%. The emission increases significantly only above 13 V and above 22V a negative differential resistance (NDR) occurs. Fig. 4a shows a spectrum for each of the dots in the IV in the inset. Each spectrum was taken adjusting the duty cycle of the current pulses in order keep a similar average dissipated power through the device. The spectra were then rescaled so that the area under the curves would correspond to the LIA signal measured in the LI curve shown in the inset. We clearly see a rapid increase in intensity starting from 16 V. The peak of the spectra remains fixed at 110 meV, confirming that this is a vertical transition. Unfortunately, after a small FWHM decrease from 65 to about 50 meV, the peak increases again because we reach the NDR. The peaks at higher energy correspond to the TM-polarized EL shown in Fig. 3b, which is not completely filtered out by the polarizer. Fig. 4b shows the temperature dependence of the TE electroluminescence. The measurements were taken under the same pumping conditions described above at a constant pump current of 4.8 A. The emission remains at 110 meV and increases at lower temperatures but it is still clearly visible at room temperature. We fitted each spectrum with multiple Lorentzian curves and then computed the integral of the Lorentzian attributed to the TE peak at 110 meV. This area, normalized to the largest area at 10K, is plotted vs. temperature in the inset in Fig. 4b. The area decreases to about 40% of its original value, which is consistent with the effects of intersubband scattering rates associated with optical phonon emission[19].

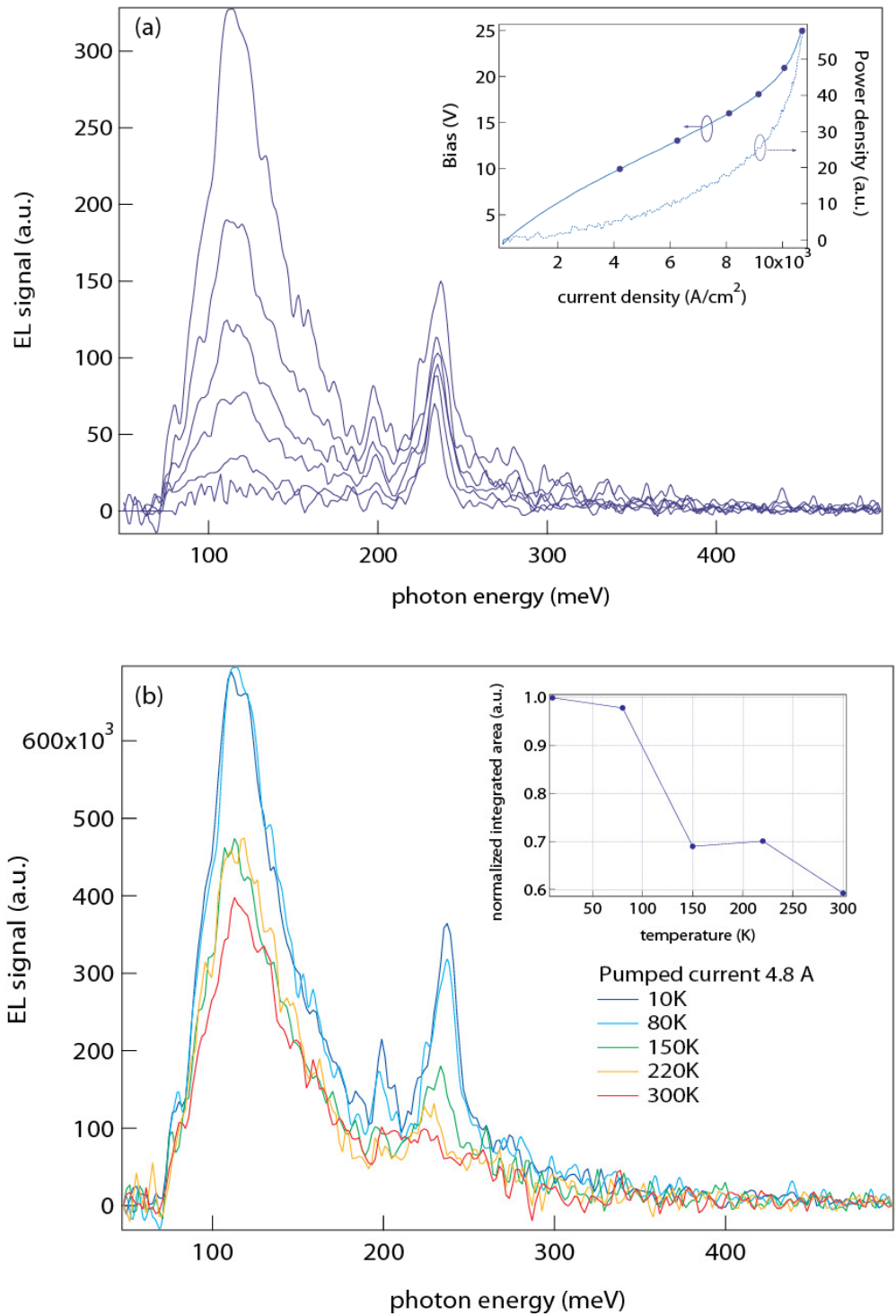

**FIG. 4** a) EL measurements at various biases for the sample exhibiting TE-polarized emission. Inset: LIV characteristics at 80K for this sample. b) Temperature dependence of the TE-polarized EL for a constant pump power of 4.8A with a 3% duty cycle. Inset: Area fitted under the TE-polarized EL curve vs. temperature.

To conclude we have demonstrated mid-infrared TE-polarized electroluminescence from intersubband transitions in a quantum cascade structure based on self-assembled InAs/AlInAs QDashes. The emission is related to a transition from the QDash first excited state, associated to the lateral confinement provided by the width of the dashes, to the QDash ground state. The lack of TE-polarized EL for ridges processed along the width of the dashes, where the electric field of the TE emission is perpendicular to the width of the dashes, is a further confirmation that the emission depends on these nanostructures. This emission is consistent with intersubband absorption measurements on a dedicated sample based only on QDashes embedded in AlInAs. Even though the EL signal increases as expected with increased pumped current, an NDR is reached at around 21 V and above this bias the FWHM of the EL increases. The TE-polarized EL is observed also at room temperature without a significant decrease compared to 80K, which is promising for the development of quantum-dash-based QCLs. We believe that an improved design of the cascade structure is necessary to engineer the population inversion needed to observe gain from QCL active regions based on these nanostructures.

This work was supported by the Swiss National Foundation under the NCCR project Quantum Photonics.